\documentclass[aps,pra,10pt,superscriptaddress,twocolumn,floatfix]{revtex4-1}

\usepackage{natbib}
\usepackage[utf8]{inputenc}
\usepackage{xcolor}
\definecolor{linkblue}{RGB}{31,119,180}
\usepackage[
  unicode,
  colorlinks,
  citecolor=blue,
  linkcolor=linkblue,
  urlcolor=linkblue,
  bookmarks=true,
  bookmarksopen=true,
  bookmarksopenlevel=3,
  bookmarksnumbered=true]{hyperref}

\usepackage{graphicx}
\usepackage{amsmath}
\usepackage{amssymb}

\begin{document}

\title{Extended analysis of distillation and purification of squeezed states of light}

\author{Jarom\'{i}r Fiur\'{a}\v{s}ek}
\affiliation{Department of Optics, {Faculty of Science}, Palack\'y University, 17. listopadu 12, 77900 Olomouc, Czech Republic}
\author{Stephan Grebien}
\affiliation{Institut f\"ur Quantenphysik \& Zentrum f\"ur Optische Quantentechnologien, Universit\"at Hamburg, Luruper Chaussee 149, 22761 Hamburg, Germany}
\author{Roman Schna\-bel}
\affiliation{Institut f\"ur Quantenphysik \& Zentrum f\"ur Optische Quantentechnologien, Universit\"at Hamburg, Luruper Chaussee 149, 22761 Hamburg, Germany}

\date{\today}

\begin{abstract}
Squeezed states of light are one of the most important fundamental resources for quantum optics, optical quantum information processing and quantum sensing. Recently, it has been experimentally demonstrated that the squeezing of single-mode squeezed vacuum states can be enhanced by probabilistic two-photon subtraction. A further enhancement of the squeezing is subsequently possible by heralded Gaussification that distills a Gaussian state from the de-Gaussified two-photon subtracted state. Here we provide an extended theoretical analysis of squeezing distillation and purification. We consider a more general scheme in which photon subtraction is combined with a weak coherent displacement. 
This more flexible scheme allows to enhance squeezing for arbitrary input squeezing value. Moreover, if the modified two-photon subtraction operation is properly chosen, then  arbitrary strong squeezing can be distilled by subsequent Gaussification.  We go beyond pure states and show that the combination of photon subtraction and heralded Gaussification cannot suppress losses that have affected the input state.  To overcome this limitation, we propose an alternative de-Gaussifying operation based on a Fock-state filter that removes the single-photon state. 
With this de-Gaussifying operation  and subsequent re-Gaussification, pure single-mode squeezed states can be distilled from a large class of mixed input states. Interestingly, we have found that squeezing distillation by two-photon subtraction is closely related to certain methods for generating Gottesman-Kitaev-Preskill (GKP) states, which are crucial for optical quantum computing.  
\end{abstract}

\maketitle

\section{Introduction}

Conditional non-Gaussian quantum operations such as single-photon addition  \cite{Zavatta2004,Barbieri2010,Kumar20013,Fadrny2024,Chen2024} and subtraction \cite{Wenger2004,Ourjoumtsev2006,Nielsen2006,Wakui2007}  represent crucial tools in modern quantum optics and optical quantum information processing. With these operations we can engineer highly non-classical non-Gaussian states from Gaussian input states \cite{Dakna1997,Dakna1999,Fiurasek2005,Ourjoumtsev2006,Nielsen2006,Wakui2007,Takahashi2008,Marek2008,Ourjoumtsev2009,Nielsen2010,Huang2015,Sychev2017,Asavanant2017,Takase2021,Chen2024,Endo2023,Eaton2019,Konno2024, Jeong2014,Morin2014,Fadrny2024} and implement various operations such as noiseless quantum amplifiers \cite{Fiurasek2009,Marek2010,Usuga2010,Zavatta2010,Park2016,Neset2024} or a nonlinear sign gate \cite{Costanzo2017}. The conditional photon subtraction is also useful for entanglement distillation. If a single photon is subtracted from each mode of a two-mode squeezed vacuum state, then one can obtain a state with increased entanglement \cite{Opatrny2000}. 
Distillation of Gaussian entangled two-mode squeezed states was experimentally realized and tested \cite{Ourjoumtsev2010,Takahashi2010,Kurochkin2014}, which has represented a major milestone in continuous-variable quantum information processing. These experiments showed that the conditional photon subtraction also distilled the squeezing of the two-mode state \cite{Dirmeier2020}.

Recently, we have experimentally demonstrated the distillation of a single-mode squeezed vacuum state by conditioning on the subtraction of two photons \cite{Grebien2022}. The squeeze factor was increased and the state de-Gaussified. The utilization of the non-Gaussian two-photon subtraction was a crucial part in the experiment, because the squeezing of Gaussian states cannot be enhanced by passive Gaussian operations and conditioning on outcomes of homodyne detection \cite{Kraus2003}. Passive Gaussian operations therefore do not allow the distillation of Gaussian squeezing, just as local Gaussian operations do not allow the distillation of Gaussian entanglement \cite{Eisert2002,Fiurasek2002,Giedke2002}. 
Specific non-Gaussian mixed input states, such as squeezed states that suffered from random phase fluctuations or random losses, can be distilled with passive Gaussian operations \cite{Heersink2006,Franzen2006}, but the amount of squeezing or entanglement \cite{Hage2008, Dong2008, Hage2010} that can be extracted in such setting is limited. Squeezed states of light \cite{Walls1983,Schnabel2017} represent an essential and irreducible resource \cite{Braunstein2005} in quantum optics and optical quantum information processing and besides quantum state engineering they find applications e.g.~in quantum sensing \cite{Polzik1992,Abadie2011,Steinlechner2013,Acernese2019} and quantum communication \cite{Furusawa1998,Bowen2003,Cerf2001,Braunstein2005RMP,Madsen2012,Gehring2015,Jacobsen2018}. Therefore, investigation of techniques to manipulate and improve squeezing of optical states is both of fundamental interest and practically relevant.

Here we present an extended theoretical study of distillation and purification of single-mode squeezed states of light by conditional photon subtraction. We go beyond the basic scheme demonstrated in our recent experiment \cite{Grebien2022} and consider two-photon subtraction combined with coherent displacement. The addition of coherent displacements \cite{Fiurasek2005} allows us to increase the squeeze factor \cite{Schnabel2017} for any input squeezed vacuum state, as well as to target arbitrary strong squeezing after the subsequent heralded Gaussification. We present an explicit optical setup that could realize this modified two-photon subtraction and discuss and optimize the success probability of this scheme. We show that the states generated by such modified two-photon subtraction have similar structure as the approximate Gottesman-Kitaev-Preskill (GKP) states \cite{Gottesman2001} which have been recently generated from two single-mode squeezed vacuum states via  photon subtraction and conditional state breeding \cite{Konno2024}. In both cases, squeezed superpositions of vacuum and two-photon number states are generated. 

Squeezing of the two-photon subtracted state can be further increased by iterative heralded Gaussification  \cite{Browne2003,Eisert2004}, which transforms the state back to a Gaussian state. Each step of iterative Gaussification requires two copies of the state that are interfered at a balanced beam splitter and one output mode is projected onto vacuum state. For initial mixed Gaussian state, the state distilled by photon subtraction  and heralded Gaussification will generally also be mixed. Interestingly, we find that if the de-Gaussifying conditional photon subtraction is replaced by a Fock state filter that removes the single-photon state, then the subsequent Gaussification converges to a pure squeezed vacuum state for a large class of noisy input states. We thus establish a protocol for simultaneous squeezing distillation and purification. For completeness, we also theoretically investigate the distillation of two-mode squeezed states by local photon subtractions from the signal and idler modes,  which can increase the squeezing of the constituent single-mode states that can be recovered if the signal and idler modes interfere at a balanced beam splitter \cite{Dirmeier2020}. While the squeezing can be enhanced in this scenario, the recovered single-mode state becomes inevitably mixed due to the residual correlations with the other mode.

The rest of the paper is organized as follows. In Section II we introduce and analyze the protocol for distillation of (single-mode) squeezed states by conditional two-photon subtraction augmented by coherent displacements. The optimization of the success probability of the squeezing distillation scheme is investigated in Section III. In section IV we compare distillation of squeezing via modified two-photon subtraction with schemes for generation of approximate GKP states. In Section V we consider distillation of two-mode squeezed vacuum states by local photon subtraction from signal and idler beams and analyze the impact of this on the squeezing of the underlying  single-mode squeezed vacuum states whose interference formed the two-mode squeezed vacuum state. In Section VI we then present a protocol for simultaneous distillation and purification of single-mode squeezing that is  based on Fock-state filter followed by iterative Gaussification. 
Finally, Section VII contains a brief summary and conclusions.

\begin{figure}[!t!]
\includegraphics[width=0.45\linewidth]{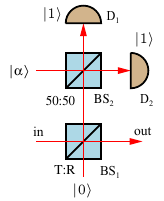}
\caption{Squeezing distillation by extended two-photon subtraction. An unbalanced beam splitter BS$_1$ with transmittance $T$ and reflectance $R$ taps off a part of the signal. The reflected beam is  interfered with an auxiliary weak coherent state $|\alpha\rangle$ at a balanced beam splitter BS$_2$. Successful squeezing distillation is heralded by detection of a single photon by each of the detectors D$_1$ and D$_2$.}
\end{figure}

\section{Squeezing distillation by two-photon subtraction}

Let $\hat{X}$ and $\hat{Y}$ denote the amplitude and phase quadrature operators. We choose to normalize them such that they satisfy the commutation relation  $[\hat{X},\hat{Y}]=2i$,
which ensures that the quadrature variances are equal to unity for vacuum and coherent states.
A pure single-mode squeezed vacuum state $|\psi(r)\rangle$ with squeeze parameter $r > 0$ exhibits a Gaussian Wigner function with anti-squeezed quadrature variance $V_X=\langle(\Delta \hat X)^2\rangle = e^{2r}$ and squeezed quadrature variance $V_Y=\langle (\Delta \hat Y)^2\rangle = e^{-2r}$, respectively, where $e^{2r} = \beta$  is the squeeze factor. The squeezed vacuum state  can be generated by acting with the unitary squeeze operator $\hat{S}(r)=\exp\left[\frac{r}{2}(\hat{a}^{\dagger2}-\hat{a}^2)\right]$  onto the vacuum state $|0\rangle$. In the Schr\"{o}dinger picture, the squeezed vacuum state can be expressed as a superposition of even number (Fock) states, 
\begin{equation}
|\psi(r)\rangle=\frac{1}{\sqrt{\cosh(r)}}\sum_{n=0}^\infty (\tanh r)^n\frac{\sqrt{(2n)!}}{2^n n!}|2n\rangle \, .
\label{eq:squeezedvacuum}
\end{equation}
We have recently shown experimentally that the squeeze factor of the state $|\psi(r)\rangle$ can be increased by conditioning on probabilistic subtraction of two photons \cite{Grebien2022}. This probabilistic transformation can be described by the operator $\hat{a}^2$, where $\hat{a}$ stands for the annihilation operator. 
In the present paper we provide a comprehensive theoretical analysis of the various aspects and generalizations of this protocol.

Since the two-photon subtraction only allows us to increase the squeeze factor of states (\ref{eq:squeezedvacuum}) if $r<1/2$ \cite{Grebien2022}, we consider here a more general setting where the photon subtraction is combined with a coherent displacement \cite{Fiurasek2005,Nielsen2010} resulting in the conditional operation $\hat{a}+\delta$. This operation can be implemented either by coherently displacing the signal before and after the photon subtraction,
\begin{equation}
\hat{D}(-\delta) \hat{a}\hat{D}(\delta)=\hat{a}+\delta \, ,
\end{equation}
or, more conveniently, by coherently displacing the tapped mode that is detected by the single photon detector \cite{Nielsen2010}. 
Here we consider a combination of two such displaced photon subtractions and choose the two coherent amplitudes such that the resulting operation preserves the parity of Fock states,
\begin{equation}
\hat{M}=(\hat{a}+\delta)(\hat{a}-\delta)=\hat{a}^2-\delta^2 \, .
\label{Moperator}
\end{equation}
In what follows we focus on the case of real $\delta^2$. Note that since $\delta$ is a complex number, $\delta^2$ can be negative as well as positive. If we set $\delta=0$ we recover as a special case the original two-photon subtraction protocol. An explicit optical scheme that implements the operation in Eq.\,(\ref{Moperator}) is shown in Fig.~1. In this section we consider the simplified idealized operation (\ref{Moperator}). The effect of the transmittance $T$ of the beam splitter BS$_1$ that performs the photon subtraction is investigated and fully taken into account in the next section.

\begin{figure}[!t!]
\includegraphics[width=0.45\linewidth]{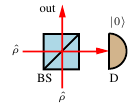}
\caption{Single step of heralded iterative Gaussification protocol \cite{Browne2003}. Two copies of the state $\hat{\rho}$ interfere at a balanced beam splitter BS and one output mode is projected  onto the vacuum state. This projection is heralded by no-click of the high-efficiency single-photon detector D. }
\label{Gaussfigure}
\end{figure}

Application of  the modified two-photon subtraction (\ref{Moperator}) to a pure squeezed vacuum state $|\psi(r)\rangle$ yields the following non-normalized state
\begin{widetext}
\begin{equation}
|\psi_{2S}(r)\rangle= \frac{1}{\sqrt{\cosh(r)}}\sum_{n=0}^\infty\left[ (2n+1)\tanh r-\delta^2\right ]  (\tanh r)^n\frac{\sqrt{(2n)!}}{2^n n!}|2n\rangle \, .
\label{psi2Sgeneralized}
\end{equation}
The variances of the amplitude and phase quadratures of this state can be expressed as
\begin{eqnarray}
V_X &=& \, e^{2r} \left[1+4 \sinh^2 r \frac{2\sinh^2 r  + \cosh r \sinh r -\delta^2}{2\sinh^4 r+(\cosh r \sinh r-\delta^2)^2}\right] ,   \nonumber \\[2mm]
V_Y &=& e^{-2r} \left[1+4 \sinh^2 r \frac{2\sinh^2 r  - \cosh r \sinh r +\delta^2}{2\sinh^4 r+(\cosh r \sinh r-\delta^2)^2}\right] \, .
\label{VXY}
\end{eqnarray}
The amplitude $\delta$ can be chosen to minimize the squeezed variance  of the two-photon subtracted state. Minimization of  $V_Y$ with respect to $\delta^2$ yields
\begin{equation}
\delta^2=\cosh r \sinh r-(2+\sqrt{6})\sinh^2 r \, .
\label{delta2optimal}
\end{equation}
For this amplitude, the quadrature variances of the two-photon subtracted state according to Eq.\,(\ref{psi2Sgeneralized}) become 
\begin{equation}
V_X =\frac{7+2\sqrt{6}}{3+\sqrt{6}} e^{2r}, \qquad V_Y =\frac{3}{3+\sqrt{6}} e^{-2r} \, , 
\label{VXYoptimal}
\end{equation}
hence the optimized two-photon subtraction increases the squeeze factor $\beta$ by the factor $\approx\!1.82$ (by $\approx\!2.6$\,dB) for arbitrary initial squeeze factors $\beta$. This procedure can generate a state with  squeeze factor at least $1.82$ from arbitrarily weakly squeezed input state, but the success probability of photon subtraction becomes small for weak initial squeezing and scales as $r^4$ for $r\ll 1$. 
The state $|\psi_{2S}(r)\rangle$ generated by modified two-photon subtraction for the squeezed vacuum state according to Eq.\,(\ref{eq:squeezedvacuum}) can be expressed as a squeezed superposition of vacuum and two-photon Fock states \cite{Fiurasek2005,Marek2008},
\begin{equation}
|\psi_{2S}(r)\rangle\propto \hat{M} \hat{S}(r)|0\rangle=\hat{S}(r)[(\cosh r\sinh r-\delta^2) |0\rangle+\sqrt{2}\sinh^2(r)|2\rangle].
\end{equation}
\end{widetext}
In particular, for the optimal $\delta^2$ given by Eq.\,(\ref{delta2optimal}) we find that the normalized state $|\psi_{2S}(r)\rangle$ reads
\begin{equation}
|\psi_{2S}(r)\rangle=\frac{1}{2\sqrt{3+\sqrt{6}}}\hat{S}(r)\left[(2+\sqrt{6})|0\rangle+\sqrt{2}|2\rangle\right]
\label{psi2Sfactorized}
\end{equation}
The state in the brackets represents the superposition of the vacuum state and the two-photon state that has the maximum quadrature squeezing among all such superpositions $c_0|0\rangle+c_2|2\rangle$. Due to the structure of the state in Eq.\,(\ref{psi2Sfactorized}), the (anti-)squeezed variances in Eq.\,(\ref{VXYoptimal}) are products of $e^{\pm 2r}$ and the fixed quadrature variances of the optimal superposition of vacuum and two-photon states.

\begin{figure*}[t]
\includegraphics[width=0.98\linewidth]{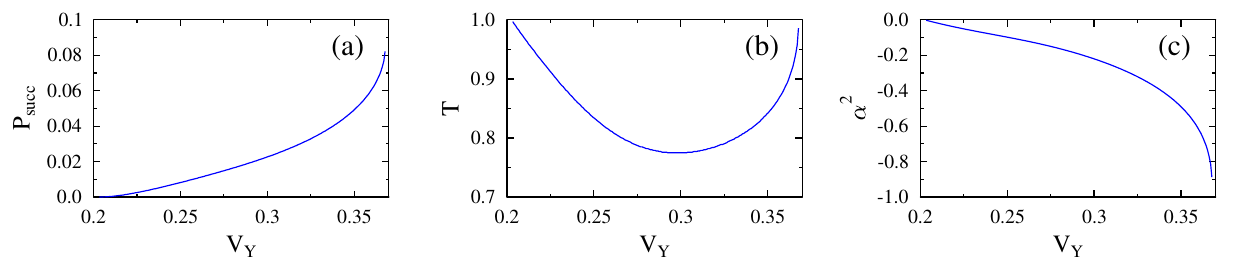}
\caption{Optimal success probability of squeezing distillation by modified two-photon subtraction for initial squeezing parameter $r=0.5$, corresponding to $V_{Y,\mathrm{ in}} \approx 0.368$. The success probability of the protocol $P_{\mathrm{succ}}$ (a), the optimal beam-splitter transmittance $T$ (b), and the optimal coherent displacement $\alpha$ (c) are plotted in dependence on the target squeezed quadrature variance $V_Y$.}
\label{figPsucc}
\end{figure*}

The squeeze factor of a photon-subtracted squeezed state can be further elevated by (iterative) heralded Gaussification \cite{Browne2003,Eisert2004,Grebien2022}. A single Gaussification step is illustrated in Fig.~\ref{Gaussfigure}. Two copies of the output state are superimposed on a balanced beam splitter, and the output mode is only accepted if the second output of the beam splitter is projected onto the vacuum state by measurement. 
The protocol can either converge to a Gaussian state or diverge. Assuming pure input states, which are superpositions of even Fock states $|2n\rangle$, the protocol preserves the ratio of amplitudes of the vacuum and two-photon Fock states.  It then directly follows from Eqs.\,(\ref{psi2Sgeneralized}) and (\ref{eq:squeezedvacuum}) that the squeeze parameter of the Gaussified state  $r_{\rm G}$ can be expressed as 
\begin{equation}
\tanh r_{\rm G}= \frac{3\tanh r -\delta^2}{\tanh r-\delta^2} \tanh r \, ,
\end{equation}
where real $\delta^2$ is assumed. 
This expression is meaningful and the Gaussification converges only if $|\tanh r_{\rm G}|<1$. Note that any required squeezing $r_{\mathrm{G}}>r$ is in principle achievable for any non-zero input squeezing $r$, if we set
\begin{equation}
\delta^2=\frac{\tanh r_{\rm G} -3\tanh r}{\tanh r_{\rm G} -\tanh r} \tanh r \, .
\end{equation}
Therefore, arbitrary strong squeezing can be distilled from arbitrary weak initial squeezing by combination of the displacement-enhanced two-photon subtraction according to Eq.\,(\ref{Moperator}) followed by iterative Gaussification.

\section{Optimization of success probability}

In this section we provide a more detailed description of the optical scheme for the modified two-photon subtraction, and we show that the parameters of this scheme can be optimized in order to maximize the success probability of squeezing distillation. The considered setup is depicted in Fig.~1. A part of the input squeezed vacuum state in mode A is reflected from an unbalanced beam splitter BS$_1$ with transmittance $T$ and reflectance $R$. The reflected beam in mode B then interferes at a balanced beam splitter BS$_2$ with coherent state $|\alpha\rangle$.   The operation succeeds when each of the  detectors D$_1$ and D$_2$ detects a single photon. In this case, the input modes B and C are projected onto the entangled two-photon state $\frac{1}{\sqrt{2}}(|2,0\rangle-|0,2\rangle)$ that is obtained by back-propagation of the two-mode Fock state $|1,1\rangle$ through BS$_2$. 
Since mode $C$ is prepared in coherent state, mode B is effectively projected onto an un-normalized state
\begin{equation}
|\omega(\alpha)\rangle_B=\frac{e^{-|\alpha|^2/2}}{2}\left(\alpha^{2}|0\rangle-\sqrt{2}|2\rangle\right) \, .
\label{stateomega}
\end{equation}
The resulting conditional operation on mode A is then the required combination of zero and two-photon subtractions combined with noiseless attenuation \cite{Micuda2012,Nunn2022,Nunn2023} of the signal imposed by the non-unit transmittance of BS$_1$,
\begin{equation}
\hat{M}_A=\frac{e^{-|\alpha|^2/2}}{2} \left(\frac{1-T}{T} \hat{a}^2 - \alpha^2  \right)t^{\hat{n}} \, .
\end{equation}
Here $t=\sqrt{T}$ is the amplitude transmittance of BS$_1$. If we set 
\begin{equation}
\alpha=\sqrt{\frac{1-T}{T}}\delta \, ,
\end{equation}
we get
\begin{equation}
\hat{M}_A=\frac{1-T}{2T} \exp\left[-\frac{1-T}{2T}|\delta|^2\right]\left( \hat{a}^2 - \delta^2  \right) t^{\hat{n}} \, .
\end{equation}
The noiseless attenuation $t^{\hat{n}}$ preserves the shape of the squeezed vacuum state according to Eq.\,(\ref{eq:squeezedvacuum}) and  it only reduces its squeeze factor to $\tilde{r}$, where
\begin{equation}
\tanh \tilde{r}=T\tanh r \, .
\end{equation}
Therefore, the quadrature variances of the state $\hat{M}_A|\psi_{SV}\rangle$ can be obtained from Eq.\,(\ref{VXY}), where one only needs to replace $r$ with $\tilde{r}$. 
The probability of success of the modified two-photon subtraction with the setup in Fig.~1 is given by
\begin{widetext}
\begin{equation}
P_{\mathrm{succ}}=  \frac{\cosh \tilde{r}}{\cosh r}  \left(\frac{1-T}{2T}\right)^2 e^{-(1-T)|\delta|^2/T}\langle \psi(\tilde{r})|(\hat{a}^{\dagger 2}-\delta^{\ast 2})(\hat{a}^2-\delta^2)|\psi(\tilde{r})\rangle \, .
\end{equation}
Assuming real $\delta^2$, this yields
\begin{equation}
P_{\mathrm{succ}}= \left(\frac{1-T}{2T}\right)^2 \frac{e^{-(1-T)|\delta|^2/T}}{\sqrt{\cosh^2 r-T^2\sinh^2 r}}\left[  (\cosh^2\tilde{r}+2\sinh^2 \tilde{r})\sinh^2 \tilde{r} -2 \delta^2 \cosh \tilde{r}\sinh \tilde{r}+\delta^4 \right] \, .
\end{equation}
\end{widetext}
For a given input state and chosen squeezed variance $V_Y$ of the output state we can optimize the parameters $T$ and $\delta$ to maximize the success probability $P_{\mathrm{succ}}$. The parameter $\delta^2$ can be determined from Eq.\,(\ref{VXY}) as a function of $T$ and $V_Y$  by solving quadratic equation, and the remaining optimization over $T$ can be performed numerically.  
As an example, we show in Fig.~\ref{figPsucc} the results of optimization of $P_{\mathrm{succ}}$ for a fixed input squeezed vacuum state with $r=0.5$ and varying target squeezed variance $V_Y$. In addition to $P_{\mathrm{succ}}$ we
display also the optimal transmittance $T$ and the squared amplitude $\alpha^2$ of the auxiliary coherent state. As the target variance $V_Y$  decreases, $T$ initially drops but then it increases again because for low output squeezed variance the reduction of squeezing by noiseless attenuation must be avoided.
The numerical results show that even an infinitesimal improvement of squeezing by the augmented two-photon subtraction incurs  a cost in terms of finite drop of the success probability below $1$.

Photon-number-resolving measurements can be perfromed with superconducting transition-edge sensors that can exhibit very high detection efficiency approaching $95\%$ \cite{Lita2008,Eaton2022}.
Nevertheless, binary on-off detectors that can only distinguish the presence or absence of photons are still utilized in most experiments. The scheme in Fig.~1 can work with such detectors provided that the transmittace $T$ is kept sufficiently high, so that the probability that more than two photons are reflected at BS$_1$ becomes negligibly small. However, if one wants to optimize the success probability of the scheme then it may not be desirable to work only in the regime $1-T \ll 1$. Approximate photon number resolution with binary on-off detectors can be achieved  by spatial or temporal multiplexing, where the detected mode is effectively split among an array of $N$ detectors \cite{Rehacek2003,Banaszek2003, Fitch2003,Achilles2003,Bartley2013,Hlousek2019,Cheng2022} and the number of detector clicks is counted.

To simplify the analysis, we have considered perfect detectors with unit detection efficiency. Non-unit detection efficiency  $\eta$ of detectors D$_1$ and D$_2$ can be modeled by two lossy channels with transmittance $\eta$ inserted just in front of the detectors. These lossy channels can be backpropagated in front of the beam splitter BS$_2$ in Fig.~1. Since coherent state remains coherent state after propagation through a lossy channel, the net effect of $\eta$ is that the mode reflected from the unbalanced beam splitter BS$_1$ propagates through a lossy channel with transmittance $\eta$ before it is projected onto the state (\ref{stateomega}).  As noted above, current superconducting single-photon detectors achieve very high detection efficiencies exceeding $90\%$ which suggests that high-quality  realization of the scheme in Fig.~1 is feasible with current technology.

\section{Relation to the generation of Gottesman-Kitaev-Preskill states}

\begin{figure}[b]
\includegraphics[width=\linewidth]{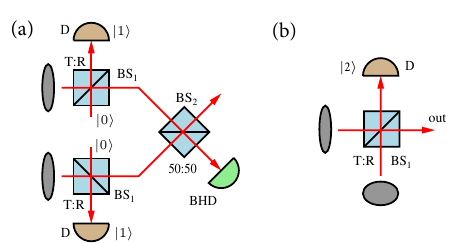}
\caption{Schemes for generation of approximate Gottesman-Kitaev-Preskill states from Gaussian squeezed states by conditional merging of two squeezed single-photon states  \cite{Konno2024} (a) or by generalized two-photon subtraction \cite{Takase2021} (b). SPD - single photon detector, BHD - balanced homodyne detector, BS - beam splitter. Ellipses illustrate the input squeezed vacuum states. }
\label{GKPfigure}
\end{figure}

\begin{figure}[!t!]
\includegraphics[width=0.6\linewidth]{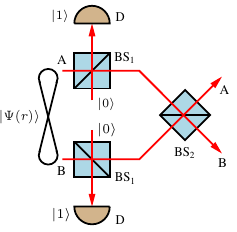}
\caption{Distillation of two-mode squeezed vacuum state. A single photon is subtracted from each mode of the  state $|\Psi(r)\rangle_{AB}$ \cite{Opatrny2000,Takahashi2010}. If the two modes A and B are subsequently interfered at a balanced beam splitter BS$_2$, then the output modes A and B exhibit enhanced  single-mode squeezing \cite{Dirmeier2020}. See text for details.}
\label{TMSVfigure}
\end{figure}

The  squeezed superpositions of vacuum and two-photon states can approximate the Gottesman-Kitaev-Preskill (GKP) states \cite{Gottesman2001} that are essential for optical quantum computing.  Very recently, such approximate GKP states of the form 
\begin{equation}
|\psi_{\mathrm{GKP}}\rangle=\hat{S}(r)(c_0|0\rangle+c_2|2\rangle) \, ,
\label{psiGKP}
\end{equation}
were generated by interference of two single-photon subtracted squeezed vacuum states at a balanced beam splitter, followed by conditioning on measurement outcomes of a balanced homodyne detector \cite{Sychev2017,Konno2024,Vasconcelos2010,Pizzimenti2024}, see Fig.~\ref{GKPfigure}(a). To show that this scheme generates the states specified in  Eq. (\ref{psiGKP}), recall that a single-photon subtracted squeezed vacuum state is fully equivalent to squeezed single-photon state, $\hat{a}\hat{S}(r)|0\rangle=\sinh(r) \hat{S}(r)|1\rangle$. Interference of two copies of this state at a balanced beam splitter yields
\begin{eqnarray}
|\Psi_{\mathrm{GKP}}\rangle_{AB}&=&\hat{U}_{\mathrm{BS}}\hat{S}_A(r)\otimes \hat{S}_B(r)|1,1\rangle \\
&=& \hat{S}_A(r)\otimes \hat{S}_B(r)\frac{1}{\sqrt{2}}(|2,0\rangle-|0,2\rangle) \, . \nonumber 
\label{PsiABGKP}
\end{eqnarray}
In the above expression $\hat{U}_{\mathrm{BS}}$ denotes the unitary operation performed by a balanced beam splitter,
\begin{equation}
\hat{U}_{\mathrm{BS}}=\exp\left[\frac{\pi}{4}(\hat{a}^\dagger\hat{b}-\hat{a}\hat{b}^\dagger)\right] \, ,
\end{equation}
and we used the fact that identical single-mode squeezing operations commute with $\hat{U}_{\mathrm{BS}}$,
$\hat{U}_{\mathrm{BS}} \hat{S}\otimes \hat{S}= \hat{S}\otimes \hat{S} \hat{U}_{\mathrm{BS}}$. Projection of the output mode $B$ onto the eigentate $|x\rangle$ of the quadrature operator $\hat{x}=\hat{X}/\sqrt{2}$ conditionally prepares mode A in squeezed superposition of vacuum and two-photon states,
\begin{equation}
_B\langle x|\Psi_{\mathrm{GKP}}\rangle_{AB}\propto\hat{S}(r)\left[(1-2e^{-2r}x^2)|0\rangle+\sqrt{2}|2\rangle\right]_A ,
\label{GKPsuperposition}
\end{equation}
 Note that the sign of the amplitude of the Fock state $|2\rangle$ in the superposition (\ref{GKPsuperposition}) can be changed by measuring the phase quadrature $\hat{Y}$ instead of the amplitude quadrature $\hat{X}$.

Alternatively, states of the form (\ref{psiGKP}) can be also prepared  by generalized two-photon subtraction, where an auxiliary single-mode squeezed vacuum state is injected into the auxiliary input port of the beam splitter that is used to subtract the two photons \cite{Takase2021,Huang2015,Tomoda2024}. This approach is illustrated in Fig.~\ref{GKPfigure}(b). As shown in Ref. \cite{Korolev2024}, the conditionally generated output state can be expressed in the $x$ representation as
\begin{equation}
\psi(x)\propto e^{- c x^2/2}(b_0+b_2 x^2)  ,
\label{psiGKPwavefunction}
\end{equation}
where the parameters $b_0$, $b_2$ and $c$ depend on the squeeze factors of the two input squeezed states and on the splitting ratio of the beam splitter $BS_1$. Since the wave function of a Fock state $|n\rangle$  is proportional to $H_n(x)e^{-x^2/2}$, where $H_n(x)$ is the Hermite polynomial of degree $n$, it is easy to see that the wave function (\ref{psiGKPwavefunction}) represents a squeezed superposition of vacuum and two-photon states. Compared to the setups depicted in Fig.~\ref{GKPfigure}, the scheme considered in the present work and depicted in Fig.~1 requires only a single copy of single-mode squeezed vacuum state and avoids conditioning on homodyne detection. 

\bigskip

\section{Distillation of two-mode squeezed states}

Consider a two-mode squeezed vacuum state
\begin{equation}
|\Psi(r)\rangle=\sqrt{1-\lambda^2}\sum_{n=0}^\infty \lambda^n |n,n\rangle_{AB} \, ,
\label{PsiTMSV}
\end{equation}
where $\lambda=\tanh r$ and $r$ is the squeeze parameter. Entanglement of this state can be enhanced by conditional subtraction of single photons from the signal and idler modes \cite{Opatrny2000,Takahashi2010,Kurochkin2014},
\begin{equation}
\hat{a}\hat{b}|\Psi(r)\rangle=\sqrt{1-\lambda^2} \sum_{n=0}^\infty (n+1)\lambda^{n+1}|n,n\rangle \, .
\label{SIsubtraction}
\end{equation}
The  two-mode squeezed vacuum state (\ref{PsiTMSV}) can be generated by combining  at a balanced beam splitter two single-mode squeezed vacuum states $|\psi(r)\rangle=\hat{S}(r)|0\rangle$ and $|\psi(-r)\rangle=\hat{S}(-r)|0\rangle$, one squeezed in the $\hat{Y}$ quadrature and the other in the $\hat{X}$ quadrature,
\begin{equation}
|\Psi(r)\rangle=\hat{U}_{BS} \hat{S}_A(r)\otimes \hat{S}_B(-r) |0,0\rangle_{AB} \, .
\label{PsiTMSVgeneration}
\end{equation}

Interestingly, the joint single-photon subtraction from signal and idler modes of two-mode squeezed vacuum state can also enhance squeezing of the constituent  single-mode states \cite{Dirmeier2020}. This enhancement of  single-mode squeezing can be revealed by letting the signal and idler modes  of the  state (\ref{SIsubtraction}) interfere at a balanced beam splitter and looking at the squeezing properties of the resulting output modes \cite{Dirmeier2020}, see Fig.~\ref{TMSVfigure}. In what follows we analyze this single-mode squeezing enhancement in more detail. 

With the help of Eq.\,(\ref{PsiTMSV}) the two-photon subtracted state given by  Eq.\,(\ref{SIsubtraction}) can be rewritten as follows,
\begin{widetext}
\begin{equation}
\hat{a}\hat{b}|\Psi(r)\rangle=\frac{1}{2}\hat{U}_{\mathrm{BS}}\left[ \hat{a}^2-\hat{b}^2\right]  \hat{S}_A(r)\otimes \hat{S}_B(-r) |0,0\rangle_{AB}= \frac{\sinh r}{\sqrt{2}}\hat{U}_{\mathrm{BS}}\hat{S}_A(r)\otimes \hat{S}_B(-r)\left[\sinh r(|2,0\rangle-|0,2\rangle)+\sqrt{2}\cosh r|0,0\rangle\right] \, . 
\label{SIUpropagated}
\end{equation}
\end{widetext}
If the signal and idler modes of the  state (\ref{SIUpropagated}) interfere at a balanced beam splitter we recover the constituent single-mode squeezed states whose squeezing was modified by the joint subtraction of two photons. If we trace over  mode B, we obtain the reduced density matrix of mode A. After normalization, we obtain
\begin{equation}
\hat{\rho}_A=\frac{1}{2\cosh(2r)}\hat{S}(r)\left [   |\varphi \rangle \langle \varphi|+ \sinh^2r |0\rangle\langle 0|  \right] \hat{S}^\dagger(r) \, .
\label{rhoA}
\end{equation}
where
\begin{equation}
|\varphi\rangle=\sqrt{2}\cosh r|0\rangle+\sinh r|2\rangle \, .
\label{varphidef}
\end{equation}
Note that the single-mode state $\hat{\rho}_A$ is mixed because the interference at a balanced beam splitter does not completely remove the correlations between the signal and idler modes of the photon subtracted two-mode squeezed vacuum. More specifically, the state $\hat{\rho}_A$ is a mixture of the Gaussian squeezed vacuum state and a non-Gaussian state obtained by squeezing the superposition of vacuum and two-photon state (\ref{varphidef}). A closed formula for the purity of the state (\ref{rhoA}) can be derived,
\begin{equation}
\mathcal{P}=1-\frac{\sinh^4 r}{2\cosh^2(2r)} \, .
\end{equation}
Variances of squeezed and anti-squeezed quadratures of $\hat{\rho}_A$ read
\begin{eqnarray}
V_X&=&e^{2r}\left[1+2\frac{e^r \sinh r}{\cosh(2r)}\right], \nonumber   \\[2mm]
V_Y&=&e^{-2r}\left[1-2\frac{e^{-r} \sinh r}{\cosh(2r)}\right] \, . 
\label{VXYtwomode}
\end{eqnarray}
We can see that the squeezing is improved for any $r$, however for large $r$ the improvement is only marginal. For weak squeezing the comparison of Eq.\,(\ref{VXYtwomode}) and Eq.\,(\ref{VXY}) with $\delta=0$ reveals that  the subtraction of two photons from a single-mode squeezed vacuum state is more efficient and leads to higher squeezing than the above discussed joint subtraction of two photons from two-mode squeezed vacuum followed by decoupling of the two modes at a beam splitter.

\section{Purification of mixed squeezed  states}

In this section, we consider distillation of mixed squeezed states. Here the goal can be two-fold:  to increase the squeezing, but also to increase the purity. The two-photon subtraction followed by Gaussification can increase the squeezing of mixed states, but it cannot suppress losses that affect the state, as we show below. Our results parallel earlier findings on limits to distillation and purification of two-mode squeezing by local de-Gaussification  and subsequent Gaussification \cite{Lund2009,Fiurasek2010}. Moreover,  we also show that an alternative de-Gaussification operation, a Fock-state filter that removes the Fock state $|1\rangle$,  together with subsequent Gaussification, can generate pure squeezed states from mixed inputs.

A mixed  single-mode squeezed Gaussian state with zero displacement is parameterized by the variances of  anti-squeezed and squeezed quadratures, $V_X$ and $V_Y$, respectively. Any mixed squeezed state with $V_Y<1$ can be obtained from some pure squeezed vacuum state with squeeze parameter $r_0$  that is transmitted through a  lossy channel $\mathcal{L}_{T_0}$ with transmittance $T_0$. The quadrature variances can then be expressed as
\begin{equation}
V_X=e^{2r_0}T_0+1-T_0, \qquad V_Y=e^{-2r_0}T_0+1-T_0 \, ,
\end{equation}
and an inversion of these formulas yields
\begin{equation}
T_0=\frac{(V_X-1)(1-V_Y)}{V_X+V_Y-2}, \qquad e^{2r_0}=\frac{V_X-1}{1-V_Y} \, .
\end{equation}

Two-photon subtraction followed by Gaussification can increase the effective squeeze parameter $r_0$, but it cannot increase the effective transmittance $T_0$. Therefore, this combination of operations cannot suppress losses in generation and distribution 
of squeezed states. To show this, observe that the operator  $\hat{M}=\hat{a}^2-\delta^2$ essentially commutes with the lossy channel $\mathcal{L}_{T_0}$. We can express the lossy channel in terms of its Kraus operators  as
\begin{equation}
\mathcal{L}_{T_0}(\hat{\rho})=\sum_{j=0}^\infty \hat{K}_j\hat{\rho}\hat{K}_j^\dagger \, ,
\end{equation}
where
\begin{equation}
\hat{K}_j=\frac{(1-T_0)^{j/2}}{\sqrt{j!}} T_0^{\hat{n}/2} \hat{a}^j \, .
\end{equation}
Since $\hat{a}^2 \hat{K}_j=T_0\hat{K}_j\hat{a}^2$ for all $j$, we have
\begin{equation}
\hat{M}\mathcal{L}_T(\hat{\psi})\hat{M}^\dagger=\mathcal{L}_T(\hat{M}_T\hat{\psi} \hat{M}_T^\dagger) \, ,
\label{Mcommutation}
\end{equation}
where $\hat{M}_T=T\hat{a}^2-\delta^2$. In particular, for $\delta=0$ we obtain
\begin{equation}
\hat{a}^2\mathcal{L}_{T_0}(\hat{\rho})\hat{a}^{\dagger 2}=T_0^2\mathcal{L}(\hat{a}^2\hat{\rho}\hat{a}^{\dagger 2}) \, .
\label{La2commutation}
\end{equation}
Therefore, two-photon subtraction applied to a mixed Gaussian squeezed state with quadrature variances $V_X$ and $V_Y$ is equivalent to subtraction of two photons from  pure squeezed vacuum state with squeeze parameter $r_0$ followed by lossy channel with transmittance $T_0$ and the same holds also for the modified two-photon subtraction $\hat{M}$.
Let $\hat{\psi}=|\psi\rangle\langle \psi|$ denote the density matrix of a pure state. Gaussification of a mixed state $\mathcal{L}_{T_0}(\hat{\psi})$ is equivalent to Gaussification of a  pure state $\hat{\psi}$ with detectors whose efficiency is reduced by factor $T_0$, followed by transmission of the Gaussified state through the lossy channel $\mathcal{L}_{T_0}$. 
Therefore, the Gaussified state will always suffer from at least the same losses as the original state.
It follows that the maximum squeezing that can be possibly extracted from a mixed squeezed state by two-photon subtraction and Gaussification is bounded by  $1-T_0$, i.e.
\begin{equation}
V_{\min} \geq \frac{V_XV_Y-1}{V_X+V_Y-2} \, .
\end{equation}

The above conclusions hold also if we consider a more realistic description of photon subtraction, where a beam splitter with transmittance $T$ is used to tap part of the signal and the reflected signal is measured with single photon detectors, c.f. Fig.~1. In this case, the two-photon subtraction is described by the operator 
\begin{equation}
\frac{1-T}{\sqrt{2}} T^{\hat{n}/2}\hat{a}^2 \, .
\end{equation}
The photon subtraction from a mixed state becomes equivalent to photon subtraction from a pure state with modified tapping beam splitter with transmittance
\begin{equation}
\tilde{T}=1-T_0(1-T) \, ,
\end{equation}
followed by lossy channel with transmittance
\begin{equation}
\tilde{T}_0=\frac{T_0 T}{T+(1-T)(1-T_0)} \, .
\label{T0formula}
\end{equation}
It follows from Eq. (\ref{T0formula}) that $\tilde{T}_0<T_0$.

As recently pointed out \cite{Zhang2024}, single-photon subtraction applied to specific mixed Gaussian states can increase their purity. Similar results can be observed also for the modified two-photon subtraction. To understand this effect, note that for suitable choice of $\delta$ the application of the operator $\hat{M}=\hat{a}^2-\delta^2$  to a  pure squeezed vacuum state can decrease the squeezing of the state, which can thus become less sensitive to losses \cite{Zhang2024}. 
If the de-Gaussified state $\hat{M}_T\hat{\psi} \hat{M}_T^\dagger$ in Eq.\,(\ref{Mcommutation}) becomes less squeezed, then it can exhibit higher purity after losses than the original  state $\mathcal{L}_T(\hat{\psi})$. However, the increased purity comes at a cost of reduced squeezing, while the goal of squeezing distillation is to increase the squeezing.

Distillation of pure squeezed vacuum states from initial mixed states is possible, however it requires a more challenging non-Gaussian operation. We found that pure-state distillation is possible with a Fock-state filter $\hat{F}_1=\hat{n}-1$
that completely eliminates the single-photon term in the density matrix \cite{Grebien2022}. The quantum filter $\hat{F}_1$ can be realised with a single-photon catalysis \cite{Ulanov2015,Lvovsky2002}, which requires an ancilla single photon state that interferes with the signal at a balanced beam splitter. Successful filtering is  heralded by detection of exactly one photon at the ancilla output port of the beam splitter. Alternatively, the operation $\hat{F}_1$ could be realized by a coherent combination of single-photon addition and subtraction \cite{Zavatta2009,Costanzo2017}, since $\hat{n}-1=2\hat{a}^\dagger\hat{a}-\hat{a}\hat{a}^\dagger$. 

For any input mixed state, the filtered density matrix $\hat{\rho}^{F}=\hat{F}_1 \hat{\rho}\hat{F}_1^\dagger$ 
will have vanishing density matrix elements $\rho_{0,1}^F$, $\rho_{1,0}^F$ and $\rho_{1,1}^F$ in Fock basis. The theory of iterative Gaussification procedure \cite{Browne2003,Eisert2004} then predicts that the Gaussification will converge to a pure  Gaussian squeezed vacuum  state whose squeeze parameter $r$ is completely determined by the parameter 
$\sigma_{2,0}^F=\rho^F_{2,0}/\rho^F_{0,0}$, namely $\tanh r=\sqrt{2}|\sigma_{2,0}^F|$. Remarkably, with the non-Gaussian operation $\hat{F}_1$ we can extract squeezing even from classical states such as coherent states. The necessary and sufficient requirement is that the initial state has non-vanishing coherence $\rho_{2,0}$ between the vacuum and two-photon Fock states, and $|\sigma_{2,0}^F|<1/\sqrt{2}$. 

It is instructive to compare purification of single-mode squeezing to the distillation and purification of two-mode squeezed entangled states. The LOCC entanglement distillation and purification protocol proposed in Ref. \cite{Fiurasek2010} involves a  nested iterative scheme, where Gaussified states have to be repeatedly de-Gaussified. 
Moreover, the de-Gaussification leading to state purification  requires two copies of the state. 
Simultaneous distillation and purification of single-mode squeezing appears to be simpler,  because the non-Gaussian operation $\hat{n}-1$ needs to be applied only once to each copy of the state and then ordinary Gaussification drives the state to pure squeezed vacuum state.

\section{Conclusions}

In summary, we have proposed and analyzed an extended scheme for distillation of single-mode squeezed states by two-photon subtraction combined with coherent displacement. The coherent displacement greatly increases the flexibility of the squeezing distillation scheme. The squeezing can be enhanced for arbitrary input single-mode squeezed vacuum state and arbitrary strong squeezing can be extracted if the photon subtraction is combined with heralded Gaussification.  Moreover, the transmittance of the beam splitter that performs the photon subtraction can be optimized to maximize the succes probability of the protocol for the chosen target squeezing. Squeezing purification and distillation of pure squeezed states is possible even from mixed input states provided that the two-photon subtraction is replaced by a Fock state filter that removes the single-photon state, and the resulting non-Gaussian state is Gaussified by heralded Gaussification.

The investigated squeezing distillation scheme based on two-photon subtraction converts squeezed vacuum state into squeezed superposition of vacuum and two-photon states. If follows that the squeezing distillation scheme is in fact  closely related to the recently demonstrated scheme for generation of approximate GKP states by interference of two squeezed single-photon states and conditioning on results of homodyne detection. Additionally, the two-photon subtracted squeezed vacuum states can approximate the even cat-like states formed by squeezed superposition of two coherent states $\hat{S}(r')(|\alpha\rangle+|-\alpha\rangle)$ \cite{Marek2008}. Merging of two squeezed single-photon states at a balanced beam splitter followed by homodyne detection of one onutput mode and conditioning on measurement outcomes close to zero can be also interpreted as cat-state breeding \cite{Sychev2017}. Therefore, the two-photon subtraction \cite{Marek2008} augmented by auxiliary coherent or squeezed states \cite{Fiurasek2005,Takase2021} provides an alternative to the breeding protocol of Ref. \cite{Sychev2017}.

The combination of single-photon subtraction and coherent displacement has already been successfully demonstrated experimentally \cite{Nielsen2010}. This, together with the recent advances in superconducting single photon detectors, suggests that the investigated scheme is experimentally feasible with current technology. In practical implementation, polarization degree of freedom can be utilized to implement the required interference between the reflected signal beam and the auxiliary coherent state \cite{Nielsen2010}, which would increase the overall stability of the setup.

\begin{acknowledgments}
S.G.~and R.S.~acknowledge support by the ERDF of the European Union and by `Fonds of the Hamburg Ministry of Science, Research, Equalities and Districts (BWFGB)'. 
J.F.~acknowledges support by the Czech Science Foundation under Grant No. 21-23120S.\\
\end{acknowledgments}

\end{document}